# CARPS: An integrated proposal and data collection system

Keith Brister, The University of Chicago

**Abstract**: Modern scripting languages and database tools combined provide a new framework for developing beam-line control and data management software. The CARPS system supports data collection by storing low level beam-line control commands in a database and playing these commands back to collect data sets. This system is combined with proposal and data management tools for support of both local and remote users.

## Introduction

Beamline control software has traditionally operated independently of the rest of the beamline support such as the proposal review and scheduling software. This is a natural outcome of the fact that beamline scientists have been responsible for the operation of the beamline while the administrative support personnel have been responsible for the proposal review and scheduling administration. Progress in both beamline controls and beamline administration has been along the lines of making electronic versions of dials and paper forms, respectively.

The CARPS project, described here, brings together the beamline controls and the beamline administration by integrating the operational and administrative databases using freely available software packages. It was developed to liberate the beamline personnel from as many routine tasks as possible.

The capabilities of the current system are:

1) Proposal submission
2) Proposals review
3) Scheduling
4) Trip administration (who and what is coming when)
5) Remote collaboration and data collection
6) Data archiving and retrieval

The proposal tools developed to date are based on use of one facility but it would be possible to extend this approach to others as the code is freely available (currently at carps.sourceforge.net).

## The Old Model

The administration and control of beamlines has changed little in concept over the past twenty years or so. In the beginning adjustments were made by hand in a time consuming manor, especially considering that the scientists could not actually touch the instruments while the x-rays were present. Naturally, motors were installed on crucial adjustments and the remote control of at least part of the beamline made possible. Simple (yet powerful) programs were developed that provided more than just beamline adjustments: plotting and state archiving were added first with coordinated motions of several motors soon following. In concept, the programs used to control scientific experiments have become automated versions what had been done by hand.

Likewise, the administration of the early facilities rapidly advanced to formal paper based proposal systems in order to make the best scientific use of scarce resources. The explosion of internet has moved many of these systems to the web, but for the most part proposals are handled in a way that differs little from what could have been done by the Phoenicians with clay tablets, albeit faster.

## The New Model

Over the past ten years or so a number of general programming tools have become freely available to manage information. These are, of course, databases and modern computer languages such as JAVA, PERL, PHP, and Python. An important characteristic of these languages is that the expertise to use them is more generally available than the knowledge of the systems used for beamline controls developed during the 80's and 90's. This has been appreciated by others and new beamline control software is being written in these new languages (see the other presentations in this session, for example).

If experiments were performed in isolation from other experiments and were completely self contained, there would be no need to go further. The existing beamline control programs and the existing proposal systems already do the job they were built for. But this is not the case: projects involve groups of people (some remote) and involve access to the data after the experiment is completed. Often several trips to several facilities are needed to bring a project to fruition

Combining the operational databases (beamline controls) and the administrative databases (proposals and user program information) has some interesting advantages: User authentication used for proposal submission and beamtime planning can now be used for remote data collection and collaboration. Project information used to communicate the experimental needs to the facility is now available at data collection time to help organize the experiment. By linking to the scheduling database, the code can tell which group may have access to the beamline controls.

## The BioCARS Implementation

Since 1998 BioCARS has used an on-line proposal system. Presently this system uses MySQL as the database manager, apache as the web server, PHP as the scripting language, and RedHat/LINUX on a DELL server. Figure 1 shows a simplified version schema currently used. The table dbUserTable contains access information for a given research group or for a staff member.

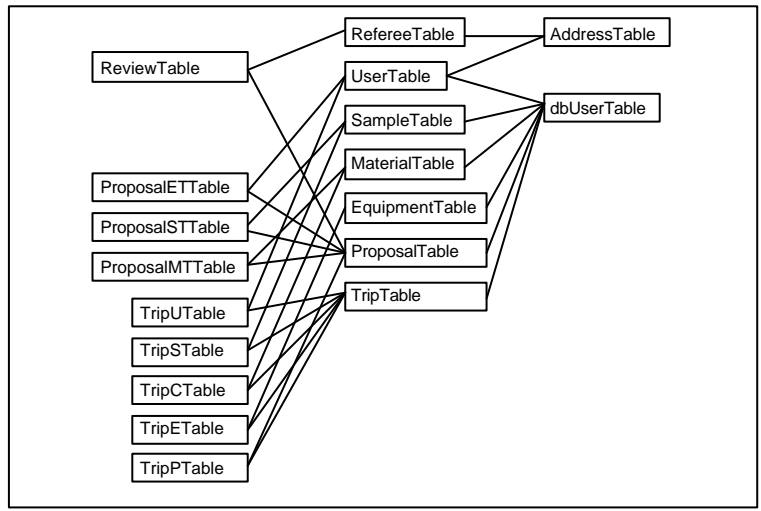

Figure 1. Simplified schema for the proposal/trip portion of the CARPS system. The left hand tables link reviews to referees and proposals, proposals to the experimental team, samples, and chemicals, and trips to the experimental team, samples, chemicals, equipment, and proposals. Tables from the operational database (not shown) also link to the trip table.

There are several features of this schema that differ from other proposal implementations:
1) The system attempts to track information by research group, not by individuals. Research groups decide for themselves how to access the system; some share the access information among all the members, having the individuals enter information. Others designate one person to enter all the information.
2) The login process involves three pieces of information: the group name, the group password, and the group's access key. The group name and password are chosen by the group to suit themselves while the access key (currently a 16 character random string), is chosen by the proposal system. An automated email sends a URL containing the access key which is subsequently stored as a cookie.
3) All information is persistent: groups can use information previously entered to generate a new proposal or a new trip.

The access key warrants a little explanation. Often groups share passwords and login information so that everyone associated with the group knows this information. When someone leaves the group they take this information with them. The access key, essentially a second password, makes it that much more difficult for someone no longer associated with the group to gain access to currently project details. It is a simple matter to change the access key without have to change the password.

The web front end main page gives user groups access to the following areas:

1) People associated with the group

2) Samples entered by the group
3) Chemicals entered by the group
4) Specialized equipment used by the group
5) Proposals
6) Trips
7) Publications

Staff members may have some of these addition areas available, depending on their role:
1) Schedule
2) Operational items, such as detector location and the type of experiment configured at a station
3) Access to other user group areas
4) Proposal preparation
5) List of users (Contact list)
6) Bulk e-mail with "mail-merge" features
7) Scheduling tools
8) Referee administration (contact information for reviewers)
9) Review administration (manage the review system)
10) Edit publication information
11) Generate "boiler-plate" information for the funding agency

## The Trip

All the sample preparation, proposal submission, and scheduling leads up to a trip to the facility. The "trip page" assembles all the details needed. The user and/or staff member selects from the lists of people, samples, chemicals, equipment, and proposals those that are relevant to this particular visit to the facility. This is where arrival, lodging, sample and chemical quantities are entered. This is also where access to data archives is given and where the access to remote collaboration and data collection tools are to be found.

## Database approach to data collection

The traditional model for beamline controls involves the low-level beamline controls and a higher level user interface. There two ways to implement this: The first is that the low-level beamline controls has all the logic to run the experiment and the user interface provides the "window dressing". EPICS with MEDM is one example of this. The second is where the low-level beamline controls are slaved to a user interface that itself contains the logic to run the experiment. The various area detector manufactures provide interfaces that act in this way.

The database approach implemented at BioCARS follows the second method: low level beamline control commands generated by a user interface are stored in a database and played back by a translator, called here the sequence engine. Figure 2 shows a schematic of how this works for exposing diffraction images (frames).

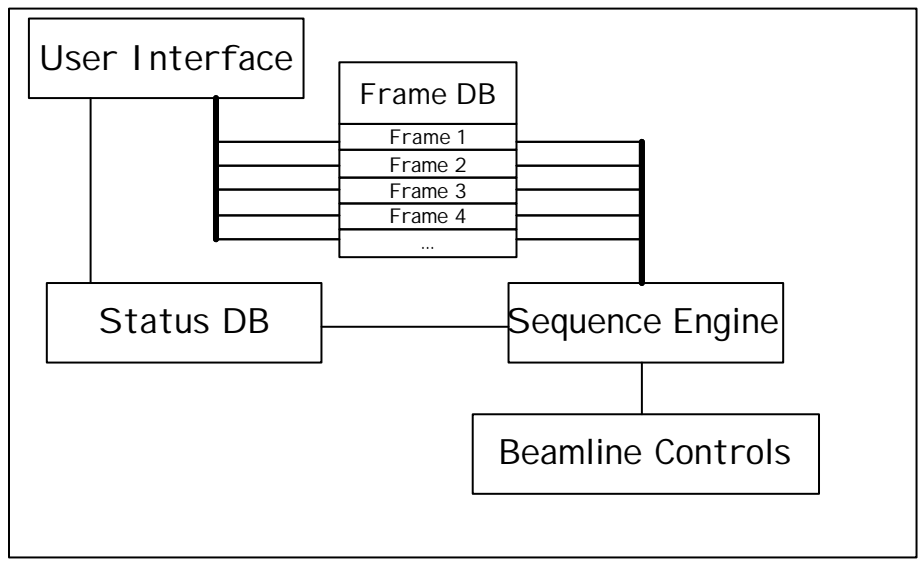

Figure 2. Schematic of the user interface-database-beamline controls interaction. The low level instructions for exposing each frame are generated by the user interface and played back by the sequence engine.

The advantage of this method, besides that of being linked to the proposal/trip database, is that the user interface is independent of both the beamline controls and the database. Any program capable of communication with the database can be used to manufacture a user interface. Multiple user interfaces may be in use simultaneously. For example, PHP is used in this implementation to generate sequences and show the status remotely while Python is used to show the status of the running experiment locally.

Programmers wanting to use this system do not need to learn the details of the communication protocol used to talk to the beamline controls; they only need concern themselves with communication with the database and this is well documented for most languages likely to be used.

## Future possibilities

The present system has been used to collect monochromatic and MAD data as well as both static and time-resolved Laue data at BioCARS (APS Sector 14). In addition, a standalone system has been set up and has been used to collect high-pressure powder diffraction data at GSECARS (APS Sector 13) as well as high-pressure single-crystal diffraction data at ChemMatCARS (APS Sector 15). Although each of these facilities uses EPICS as the low-level beamline controls, the only part of the system that cares is the sequence engine. The sequence engine itself is not a major project at all; sequence engines for other low-level beamline control systems should not be difficult to write.

In fact, the sequence engine does not need to be limited to communications with beamline control systems. There would be an advantage to linking to data analysis packages to at least start up the analysis of the data.

Likewise, the database does not need to be limited to use by the facility. Users could bring to the facility their own interface to the experiment. This interface could be integrated with the user's research database allowing management of data collected at several beamlines for the same project.

In an effort to facilitate others in using this package, the project has been moved to carps.sourceforge.net under the Gnu Public License (GPL).